\documentclass[aps,prd,twocolumn,groupedaddress,superscriptaddress]{revtex4-2}
\pdfoutput=1 
\usepackage{graphicx}  
\usepackage{subfigure} 
\usepackage{bm}        
\usepackage{amsmath}   
\usepackage{cancel} 
\usepackage{verbatim} 
\usepackage{hyperref}
\usepackage{seqsplit}

\begin{document}

\title{Versatile Multi-MW Proton Facility with \\ Synchrotron Upgrade of Fermilab Proton Complex}
\author{J.~Eldred, R.~Ainsworth, Y.~Alexahin, C.~Bhat, S.~Chattopadhyay, P.~Derwent, D.~Johnson, C.~Johnstone, J.~Johnstone, I.~Kourbanis, V.~Lebedev, S.~Nagaitsev, W.~Pellico, E.~Pozdeyev, V.~Shiltsev, M.~Syphers, C.Y.~Tan, A.~Valishev, and R.~Zwaska}
\affiliation{Fermi National Accelerator Laboratory, Batavia, Illinois 60510, USA}


\maketitle

\section{Introduction}


DUNE/LBNF constitutes an international multi-decadal physics program for leading-edge neutrino science and proton decay studies~\cite{DUNE} and is expected to serve as the flagship particle experiment based at Fermilab.

\begin{figure}[htp]
\begin{centering}
\includegraphics[height=120pt]{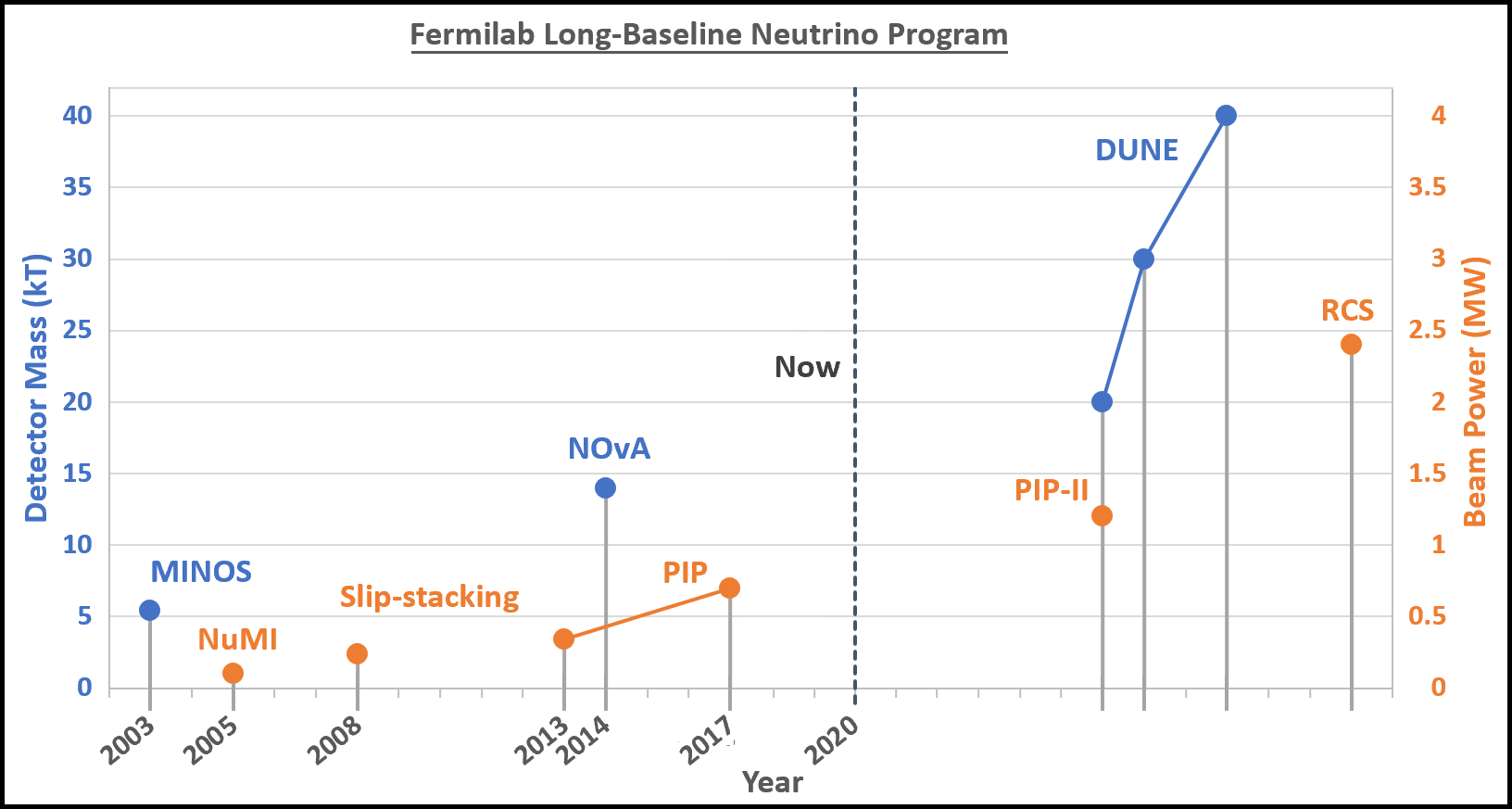}
 \caption{Past and projected milestones in Fermilab long-baseline neutrino program, as measured in detector mass and 120~GeV beam power at the Main Injector.~\cite{Eldred}} 
  \label{History}
\end{centering}
\end{figure}

Whenever Fermilab has advanced the scale of its long-baseline neutrino detectors, it has been advantageous to increase proton power to the neutrino source commensurately. Fig.~\ref{History} shows a timeline for detector and accelerator milestones of the Fermilab long-baseline neutrino program.

The Fermilab Main Injector (MI) is expected to provide 1.2~MW at 120~GeV for the DUNE/LBNF program, with the PIP-II upgrade~\cite{PIP2}. However the DUNE/LBNF science program also anticipates an upgrade of the Fermilab proton complex to 2.4~MW at 120~GeV in the Main Injector. With the 2.4-MW upgrade, DUNE is competitive with and complementary to other long-baseline neutrino experiments proposed on a similar timescale~\cite{ICFA,T2K,ESS}. The upgrade to 2.4~MW beam power for a 120-GeV MI cycle should also enable at least 2.15~MW for a 80-GeV MI cycle and 2.0~MW for a 60-GeV MI cycle.



A scenario of achieving 2-MW beam power in the Fermilab Main Injector by replacing the Booster with a new rapid-cycling synchrotron (RCS) was originally laid out in the 2003 Proton Driver Study II (PD2)~\cite{PDriver}. In 2010, a superseding RCS proposal was laid out in the Project X Initial Configuration Document 2 (ICD-2)~\cite{ICD2}. In its modern incarnation~\cite{Nagaitsev}, the ICD-2 proposal uses a 2-GeV upgrade of the PIP-II linac, features a cost-effective 8-GeV RCS which ramps at 10~Hz, accumulates batches (pulses) in the Recycler, and achieves a 2-MW Main Injector. A separate RCS scenario for a 2.4-MW Main Injector, featuring a 1-GeV linac, 15-Hz 11-GeV RCS, and slip-stacking was considered in \cite{Eldred,Eldred2}.


There are a range of self-consistent RCS upgrade scenarios compatible with a 2.4-MW upgrade of LBNF/DUNE, and with technical challenges well-considered. However the RCS scenarios differ in their implications for beamlines at other energies, performance requirements for the Main Injector, and compatibility with further power upgrades. To refine the baseline RCS upgrade path for a powerful science-driven proton complex at Fermilab, the Fermilab Booster Replacement Committee was formed with a science working group and an accelerator working group in close collaboration.


In particular, the Booster Replacement Science Workshop~\cite{BRworkshop} was held on May 19 2020 to explore the physics potential enabled by the Booster Replacement upgrade. Subsequently, an RCS Task Force was created at Fermilab to construct a unified RCS scenario that drives the ambitious science program for DUNE/LBNF, supports a variety of compelling experiments on other beamlines, and enables futures multi-MW power upgrades.

This work is ongoing, we provide an early outline here.


\section{High-Power LBNF \& Main Injector}

The Main Injector must be able to accommodate the 2.4~MW beam operation and should also be upgradeable to higher beam power~\cite{Tau}. For Main Injector reliability and upgradeability, a 2.4-MW upgrade that does not rely on slip-stacking or the Fermilab Recycler ring could be considered. Limiting the Main Injector intensity to 200e12 protons would also facilitate target design for the 2.4~MW LBNF program (see also \cite{target}). That intensity limit would also be compatible with a later upgrade path to 4~MW Main Injector beam power by upgrading the RF and magnet power to reduce the Main Injector ramp rate~\cite{Kourbanis}. 


For an 8~GeV RCS beam, much of the existing Booster-MI transfer line infrastructure can be re-purposed. However by constructing a new 12~GeV transfer line to MI-10, the Main Injector space-charge tune-spread and geometric emittance need not exceed that of PIP-II era Main Injector operation.

An RCS with circumference up to 570~m would be capable of accumulating up to five batches (pulses) in the Main Injector. To limit the Main Injector intensity below 200e12 protons while also achieving the 2.4~MW benchmark at 120~GeV and 2.0~MW benchmark at 60~GeV, the RCS must ramp at least 20~Hz (without a stacking ring). For a 120~GeV the Main Injector cycle time is 1.4~s, determined by 1.2~s for the Main Injector ramp and (5-1)/20 s to fill the Main Injector.



\section{Science Program at RCS Beamline}

The direct extraction beamline of the RCS should provide at least 0.5~MW beam power with energies in the range 8-16 GeV to support a leading-edge pulsed power program. Supported experiments include a kaon decay-at-rest program~\cite{KDAR}, dark matter search from intermediate energy protons~\cite{DM10}, a proton irradiation facility~\cite{pRad}, and any successor experiments to the current short-baseline neutrino program~\cite{SBND}. In the scenario outlined in the section below, the RCS provides 1.15~MW power at 12~GeV concurrently with 2.4~MW MI operation, with an intensity of 37e12 protons and ramp rate of 20~Hz.

At about 2~MW, the RCS beamline would be able to serve as a frontend for a neutrino factory and associated muon collider R\&D program~\cite{MuonC}. The RCS can be designed to upgraded to a 30~Hz ramp rate; in the scenario outlined below this achieves 2.1~MW at 12~GeV (using all RCS cycles).

\section{RCS Scenario}

The RCS parameters described in Table~\ref{Param} fulfill the requirements described in the previous two sections, to enable 2.4~MW Main Injector operation and MW-class 12~GeV beamline program. The Main Injector program is upgradeable to 4~MW and the 12-GeV beamline program to 2~MW (but not both concurrently).

\begin{table}[htp]
\begin{tabular}{|| l | l |}
\hline
Parameter & Value \\
\hline
RCS Intensity & 37 e12 \\
RCS Circumference & 570~m \\
Number of RCS batches & 5~batches \\
RCS Rep. Rate & 20~Hz \\
RCS Norm. Emit. (95\%) & 24~mm~mrad \\
RCS Injection Energy & 2~GeV \\
RCS Extraction Energy & 12~GeV \\
Avail. 12~GeV Power (120~GeV MI cycle) & 1.15~MW \\
\hline
MI Intensity & 185 e12 \\
MI Cycle Time (120-GeV MI cycle) & 1.4~s \\
MI Power (120-GeV MI cycle) & 2.4~MW \\
\hline
\end{tabular}
\caption{Achievable RCS parameters that fulfill power requirements for ambitious multi-faceted science program at RCS beamline and DUNE/LBNF.}
\label{Param}
\end{table}



At 0.8~GeV the RCS would have an extreme space-charge tune-shift of -0.64, but the tune-shift can be suppressed down to -0.2 by upgrading the PIP-II linac energy to 2~GeV. The RCS lattice design should also be superperiodic to enhance dynamic aperture. A superperiodic RCS lattice design would facilitate an RCS design which can (optionally) be made compatible with integrable optics or electron-lens technology~\cite{IOTA1,IOTA2,EldredI}.




To minimize uncontrolled losses and emittance growth, the RCS would feature phase-space painted injection, modest space-charge, aggressive collimation, and transition-free acceleration. The RCS can use a metalized ceramic beampipe to prevent eddy-current heating. 

\section{2 GeV Pulsed Proton Program}

A 3~ms injection time is required to fill the RCS to 37e12~protons with the 2~mA PIP-II linac beam, which presents two challenges. The first challenge is that the bend field changes by 1-2\% over the course of the injection time due to the 20-30~Hz resonant-circuit ramping magnets. The second challenge is foil-stripping injection for 3~ms at high-energy, which requires a long injection straight, large beta functions and high-power beam collimators to control foil temperature and beam scattering. 

Retrofitting the PIP-II linac as a 5-10~mA pulsed linac would alleviate both of these challenges with the multi-ms fill time, but eliminate the possibility of a linac-based physics program. The proposed linac-based physics program includes a mu2e-like charged-lepton flavor violation experiment~\cite{mu2e2,CLFV}, low energy muon experiments~\cite{LowMu}, and the REDTOP run-II/run-III program~\cite{REDTOP}.

Alternatively, a 2~GeV storage ring would also alleviate both challenges associated with the multi-ms fill time, but instead expand the range of science programs that can be accommodated. The 2~GeV storage ring could use permanent or DC-powered magnets, share the same tunnel with the RCS, and have wider apertures and longer straight sections. The 2~mA beam could be foil-injected using several 120~Hz painting cycles and then transferred to the RCS for immediate acceleration. 

Although foil-stripping technology would be sufficient to fulfill the requirements of the RCS program, the development of an emerging laser-stripping technology~\cite{Cousineau} may potentially allow for MW-class beam power to simultaneously be available for a 2~GeV pulsed proton program. That new 2~GeV program would be comparable in capability to a spallation neutron facility but could be designed to serve a new particle physics program. The proposed science program could include stopped pion source experiments~\cite{DM1}, PRISM-like charged-lepton flavor violation experiments~\cite{CLFV}, and/or neutron-antineutron oscillation experiments~\cite{NNbar1,NNbar2}. Several proposed experiments require high-power with short beam pulses, making methods of pulse compression an important design question.

\end{document}